\documentstyle{prhep97}

\makeatletter
\let\chapter\hid@chapter
\makeatother
\begin{document}


\authorrunning{D. Atwood and A.\,Soni}
\titlerunning{{\talknumber}: Penguins, Charmless $B$ Decays}
 

\def\talknumber{901} 

\title{{\talknumber}: Penguins, charmless $B$ decays and the hunt for CP
violation}
\author{Atwood, David\inst{1} (atwood@iastate.edu) \\
Amarjit\,Soni\inst{2} (soni@bnl.gov)}
\institute{\llap{$^1$}\vspace*{-11pt}Dept. of Physics and Astronomy, Iowa State Univ., Ames, IA\
\ 50010 \\
\llap{$^2$}Brookhaven National Laboratory, Upton, NY\ \ 11973; (presenter)}

\maketitle

%
\section{Introduction}

By a multitude of very important observations involving the QCD
penguin, CLEO \cite{miller} has declared 1997 to be the year of the Strong
Penguin. 
The observed size of the various modes suggest that the penguin is
rather robust. The experimental observations have fueled intense
theoretical activity some of which we will briefly review here. For
convenience, we have divided the theoretical activity into four
categories (See Table~\ref{tabone}): Exclusive modes, inclusive
$\eta^\prime$ modes, the charm deficit and implications for CP
violation. 

\begin{table}
\caption{Sample of Recent Theoretical Works\label{tabone}}
\begin{center}
\begin{tabular}{lcl}
{\bf Who} & {\bf What} & {\bf Comments} \\ \\
Ciuchini {\it et al}. \cite{ciuchini} & 2 Body Modes & Charming Penguins \\
Ali \& Greub \cite{ali} & 2 Body Modes & Improve Factorization,
$\eta_c\leftrightarrow \eta^\prime\cdots$ \\
Cheng \& Tseng \cite{cheng} & $\eta^\prime(\eta) +K(K^\ast, \rho,\pi)$ \quad & Improve  
Factorization, $\eta_c\leftrightarrow \eta^\prime \cdots$ \\
Kagan \& Petrov \cite{kagan} & $\eta^\prime K$ & SM Factorization OK$\cdots$ \\
Datta {\it et al}. \cite{datta} & $\eta^\prime K(K^\ast)$ & Factorization and
Anomaly \\
Shuryak \& Zhitnitsky \cite{shuryak} & $\eta^\prime K$ & $\eta_c\leftrightarrow
\eta^\prime \cdots$ \\
Halperin \& Zhitnitsky \cite{halperin} & $\eta^\prime K$ & $\eta_c\leftrightarrow
\eta^\prime\cdots$ \\
\hline
\\
Halperin \& Zhitnitsky \cite{halperin} & $\eta^\prime X_s$ & $\eta_c\leftrightarrow
\eta^\prime\cdots$ \\
Hou \& Tseng \cite{hou} & $\eta^\prime X_s$ & New Physics \\
Kagan \& Petrov \cite{kagan} & $\eta^\prime X_s$ & New Physics \\
Yuan \& Chao \cite{yuan} & $\eta^\prime X_s$ & $(\bar cc)_8\to \eta^\prime X_s$ \\
Datta {\it et al}. \cite{datta} & $\eta^\prime X_s$ & Factorization${}+{}$Anomaly
\\
Atwood \& S \cite{atwood} & $\eta^\prime X_s, \eta X_s\cdots$ & QCD Anomaly \\
\hline \\
Dunietz {\it et al}. \cite{dunietz} & Charm Deficit \\
Lenz {\it et al}. \cite{lenz} & Charm Deficit \\
\hline \\
Fleischer \& Mannel \cite{fleischer} & $K\pi$ & $\gamma$ \\
London \& S \cite{london} & $\eta^\prime K+\cdots$ & $\beta_{penguin}$ \\
Dighe {\it et al}. \cite{dighe} & $\eta^\prime(\eta) + K(\pi)$ & Direct CP \\
Hou \& Tseng  \cite{hou} & $\eta^\prime X_s$ & Non-Std-CP \\
Kagan \& Petrov  \cite{kagan} & $\eta^\prime X_s$ & Non-Std-CP \\
Atwood \& S  \cite{atwoodtwo} & $\eta^\prime X_s\cdots$ & Non-Std-CP 
\end{tabular}
\end{center}
\end{table}

\section{The challenge of pure hadronic (exclusive) modes}

There are renewed attempts to understand $B$-decays to two-body
hadronic modes involving the penguins. Most of the works are centered
around modifying factorization. The starting point in these
calculations is the next-to-leading-order (NLO) short-distance (SD)
Hamiltonian which at the quark level is clearly rather precise with a
minimal uncertainty due to scale. However, it is quite unclear as to
the advantage of using the NLO apparatus as the calculations of the
hadronic matrix elements is highly uncertain. See below.

\section{Charming-Penguins}

Martinelli {\it et al}.\ \cite{ciuchini} have made the interesting observation
that 
graphs containing $c\bar c$ loops (the so-called ``charming penguins'')
could be important for final states involving the light states (such as
$K\pi$, $\pi\pi\cdots$) and therefore should not be ignored as they are
commonly done in calculation of matrix elements. The idea of charming
penguin is closely related to the eye-graphs which are believed to be
important for the emergence of the observed $\Delta I=1/2$ rule in
$K$-decays. Of course, their role in $B$-decays is not known.

\section{$\eta_c$-$\eta^\prime$ mixing \protect\cite{berkel,ali,cheng,shuryak,halperin,yuan}}

Since the decay $b\to c\bar cs$ is accompanied by a hefty CKM factor,
$\eta_c\leftrightarrow \eta^\prime$ mixing 
could 
possibly 
become important
in $B\to \eta^\prime$ decays. However, it is quite tricky to isolate this
mixing as such. For one thing, some of the glue in the $b\to sg^\ast$
originates from $c\bar c$ annihilation via $b\to c\bar cs$. So one may
equally well think of the $\eta^\prime$ originating from this glue. More
specifically mixing with $\eta_c$ and with 2-glue are interrelated. Quite
understandably there is an enormous variation in the estimated
$\eta_c\leftrightarrow \eta^\prime$ mixing.

Table~\ref{tabtwo} shows a comparison of some of the {\it recent\/}
calculations of exclusive 2-body modes. Notice that $\eta^\prime
K^\ast$ is an excellent discriminator amongst various models. The $\pi^0\pi^0$
mode which is 
especially important for $\alpha$-extraction appears most difficult to
pin down; almost all the models seem to find it below $10^{-6}$ but the
predictions are not at all reliable.

\begin{table}
\caption{ Recent studies of exclusive modes; Br's 
in units of $10^{-5}$\label{tabtwo}}
\begin{center}
\begin{tabular}{l|c|c|c|c|c}
\hline
&\multicolumn{4}{|c|}{Sample Studies} & \\ \cline{2-5}
Mode & AG \cite{ali} & CT \cite{yuan} & KP \cite{kagan} & Romans \cite{ciuchini}
& CLEO \cite{miller} \\ \hline
$K^+\pi^-$ & 1--3 & & & & $1.5^{+.5}_{-.4} \pm.1\pm.1$ \\
$K^0\pi^+$ & & & 1--2 & & $2.3^{+1.1}_-{1.0}\pm.2\pm.2$ \\
$\pi^+\pi^-$ & .4--2.4 & & & & $<1.5$ \\
$\pi^0\pi^0$ & .02--.08 & & & (.05--.1)$^\ast$ & \\
$\eta^\prime K^\pm$ & 5--6 & 6--7 & 1--12 & &
$7.8^{+2.7}_{-2.2}\pm1.0$ \\
$\eta^\prime K^{\ast\pm}$ & \underbar{.07--.16} & \underbar{1--2} & &
\fbox{6--9} & $<29$ \\
$\eta K^\pm$ & .01--.04 & .2--.5 & .1--.5 & .01--.5 & $<0.8$ \\
$\eta K^{\ast\pm}$ & .1--.2 & .3--.8 & & .1--4 & $<24$ \\ \hline
\end{tabular}
\end{center}
\end{table}

All of these theoretical calculations are extremely uncertain and have
a huge range. In large part this is a reflection of the fact that
scores of assumptions and approximations have been made to arrive
at these numbers. It is perhaps useful to list some of the assumptions
and approximations typically used in these calculations:

\begin{enumerate}

\item Effective Wilson coefficients  
are same for $B\to D\pi$ and $B\to K\pi$.

\item Color suppression works as well for 
($\pi\pi$, $K\pi\cdots$) as for ($D\pi$,
$D_sK\cdots$). 

\item Color suppression works as well for matrix elements of penguin
operators as well as it does for matrix elements of tree operators.

\item Factorization works just as well for penguin operators as 
for tree ones.

\item The eye-graph with the $c$-loop (i.e.\ the charming penguin) does
not contribute to light final states (such as $K\pi$, $\pi\pi\cdots$)
even though lattice studies over the years have suggested that such 
graphs are important for the emergence of the $\Delta I=1/2$ rule in
$K\to\pi\pi$ decays.

\item Annihilation or exchange contributions are neglected even though
these are intimately related to ``factorizable
contributions" via FSI. 

\item Penguin matrix elements are extremely sensitive to
the numerical value of current quark mass, $m_s$. 

\end{enumerate}

Due to this very long list of assumptions and approximations the
calculations for hadronic decays are highly unreliable. Thus deviations
of the experimental numbers for the absolute rates from the crude
theoretical estimates that are available cannot be used as a reliable
hint for the presence of new physics. Search for CP violating
asymmetries in some of the modes can be a much more reliable test of
new physics.

\section{A possible faint silver lining}

Despite the morass of dealing with pure hadronic modes there is perhaps
sign of a silver lining. The point is that $B$ decays have a multitude
of light-light (2-body) final states. The theoretical information that
goes into these calculations is highly correlated. So despite the
plethora of assumptions, theoretical models can easily run into
trouble. For example, it is difficult to get a hierarchy $Br(B^+ \to
K^0 \pi^+)>Br (B^0\to K^+\pi^-)> Br (B^0\to \pi^+\pi^-)$. This is
especially true for $K^0\pi^+$ versus $K^+\pi^-$. So if improved
experiments confirm the present trend then it would provide 
useful constraint on the models. The modes $wK$,
$\eta^\prime K^\ast$, $\eta K$, $\eta K^\ast$ can also be very useful
tests of models.

\section{$B\to \eta^\prime + X_s$}

Some of the
important issues are: SM vs.\ new physics, the form factor for
$g^\ast\to\eta^\prime g$, the $c\bar c$ content of the $\eta^\prime$,
direct CP violation etc.

We have made a specific proposal that a large fraction of the inclusive
signal $B\to \eta^\prime +X_s$ originates from the penguin graph
through the fragmentation
$g^\ast\to g\eta^\prime$ via the QCD anomaly \cite{atwood}. The form factor
$[H(q^2_1, q^2_1, m^2_{\eta^\prime})]$ at the anomalous vertex was
estimated by using the 
measured rate for $\psi\to\gamma\eta^\prime$\cite{atwood}.
Explicit calculations indicate that $\psi\to \gamma\eta^\prime$ is
dominated by near ``on-shell'' gluons, i.e.\ $q^2_1\sim q^2_2 \sim0$.
Since the gluon in $b\to sg^\ast$ is typically with $q^2_1\sim 5$--10
GeV$^2$, $q^2_1$ dependence of $H$ becomes quite important.

While we are not aware of any theoretical study on the dependence of
$H$ on $q^2_1$ for $g^\ast \to\eta^\prime g$, the corresponding case of
the QED anomaly for $\pi^0\to\gamma^\ast +\gamma$ has received some
theoretical attention \cite{brodsky}. Although the details vary there is broad
agreement as to how the form factor for $\pi^0\to\gamma^\ast\gamma$ due to
the quark loop scales. If one assumes that $g^\ast\to \eta^\prime
g$ form factor is essentially the same as $\gamma^\ast\to \pi^0\gamma$
then the anomaly contribution to $B\to \eta^\prime X_s$ would become
negligible. However, there are at least two reasons to think that
$g^\ast\to \eta^\prime -g$ effective form factor is quite different
from $\gamma^\ast\to \pi^0-\gamma$.

Interactions of $\eta^\prime$ with a gluon are likely to be
significantly different from that of the $\pi^0$ with $\gamma$.
Specifically, for the former case the interaction does not have to
proceed through a $\bar qq$ loop. Given that the $\eta^\prime$ owes its
existence to the gluons, the $\eta^\prime$ wave
function should contain $G\cdot \tilde G$. 
The important dimensionful parameter for the
$g^\ast$-$\eta^\prime$-$g$ form factor may not be $f_\pi$ (or
$f_{\eta^\prime}\sim f_\pi$) but rather the effective gluon mass,
$m^{eff}_g$. Lattice calculations as well as phenomenological arguments
suggest $m^{eff}_g\sim 500$--700 MeV\null. Since $(m^{eff}_g/f_\pi)^2
\sim16$--25 the anomaly contribution to $B\to \eta^\prime +X_s$ will be
significantly more than estimated by the use of the pionic form factor.

Furthermore, the $g^\ast$-$\eta^\prime$-$g$ form factor may also be
greatly influenced by the presence of nearby gluonia or states rich in
gluonic content. The gluon from the penguin can combine with a soft
glue to make such a state which subsequently decays to
$\eta^\prime+{}$light hadrons. These states can be
searched  by the resonant structure in $\eta^\prime+\pi\pi$,
$\eta^\prime+K\bar K\cdots$ amongst the $\eta^\prime+X_s$ events.

\subsection{Enhanced $b\to sg$ via new physics}

Hou and Tseng and Kagan and Petrov suggest that the large $B\to
\eta^\prime X_s$ signal is due to a very large branching ratio for
$b\to sg$ due to new physics: i.e.\ about 100 times the SM value. It is
difficult to see how such a major perturbation from new physics would
not affect $B\to X_s\gamma$; after all gluon emission enters this
calculation in important ways \cite{alitwo}.

Also Ref.\cite{kagan} argues that the 
exclusive $B\to \eta^\prime K$ signal is OK
for the SM and the inclusive is problematic. Since the
inclusive/exclusive ratio is about $8\pm2$ and is in the same ball-park
as for $(B\to\gamma X_s)/(B\to \gamma K^\ast)\sim$~~~ it is difficult
to appreciate their concern. 

\subsection{Enhanced $b\to sg^\ast$ in the SM}

An important point to note is that $b\to sg^\ast$ may be significantly
enhanced over the expectation of perturbation theory due especially to
the possibility of enhanced FSI in the dominant decay, $b\to c\bar cs$.

\section{Correspondence with $\psi$ decays}

Examination of $\psi$ decays reveals several interesting final states
which presumably result from fusion of two gluons. In general one
expects these states to have $J^{PC} = 0^{++}$, $0^{-+}$,
$2^{++}\cdots$ etc. 
Many of these states appear in radiative $\psi$ decays
with branching ratios comparable to $\psi\to\gamma\eta^\prime$. In
particular, there appears to be a close correspondence between $G\cdot
\tilde G$ (i.e.\ $0^{-+}$) and $G\cdot G$ (i.e.\ $0^{++}$) 
\cite{atwoodtwo}. 
So we should expect $f_0$, 
with appreciable BR's as well.

\section{Non-Standard CP}

As is well known in $b\to s$ transitions, CP violation effect due to
the SM are expected to be suppressed as the relevant CKM phase
$\sim0(\eta\lambda^2)$. 
Due to their big rates they can be poweful in
searching for non-standard 
CP phase.

For simplicity we can assume that $b\to s$ penguin has no SM CP-odd
phase, but due to the $u\bar u$, $c\bar c$ threshold the SM
contribution possesses a CP-conserving strong phase. Non-SM
interactions possessing a CP-odd phase may contribute another
amplitude. Interference between the two leads to direct CP which can
cause partial rate asymmetry in e.g. $B \to\eta^\prime X_s$.


Three recent studies have examined such asymmetries. In the HT \cite{hou}
and KP \cite{kagan}
works, the non-standard phase is assumed to be in the chromomagnetic
form factor for $b\to sg$. Also, as mentioned before,  they assume that the
rate for $b\to sg$ 
is enhanced over the SM by about two orders of magnitude. In our work
\cite{atwoodtwo} 
we allow the non-standard CP-odd phase to reside in the chromo-electric
or chromo-magnetic form factor. In fact the chromo-electric form factor
is found to lead to larger asymmetries. Our study also showed that
appreciable asymmetry (8--18\%) can arise even if Non-Standard-Physics
contributes only 10\% to the production rate. This is especially
significant as comparison of the rate between experiment and theory is
extremely unlikely to show the presence of such a source of new physics
and yet CP search can prove to be a viable thermometer. We also
emphasize that there are many final states with similar asymmetries. A 
specially interesting mode is $B\to
K\bar KX_s$ i.e.\ with three kaons \cite{atwoodtwo}.

\section{Summary}

CLEO has seen signals of a rather robust QCD penguin in several
exclusive channels providing new impetus for improved theoretical
understanding. The observed modes so far do not seem to require (a
large) intervention by new physics. Many of them are useful 
to search for a non-standard CP 
phase. Page limits forced many omissions, e.g. $\gamma$\cite 
{fleischer}.

\section{Acknowledgements}

We thank Ahmed
Ali, Karl Berkelman, Tom Browder, Hai-Yang Cheng, Robert Fleischer, Christopher Greub, 
Michael Gronau, Laura Reina and Jim Smith. This research was
supported under DOE contracts  DE-AC02-76CH00016 (BNL) and
DE-FG02-94ER40817 (ISU).  

%
%


\begin{thebibliography}{99}

\bibitem{miller} See D. Miller [CLEO] talk in these proceedings.

\bibitem{ciuchini} M. Ciuchini {\it et al}., hep-ph/9703353;
hep-ph/9708222; see also A. Buras and R. Fleischer, PLB341,379 (1995)  

\bibitem{ali} A. Ali and C. Greub, hep-ph/9707251.

\bibitem{cheng} H.-Y. Cheng and B. Tseng, hep-ph/9707316.

\bibitem{kagan} A. Kagan and A. Petrov, hep-ph/9707354.

\bibitem{datta} A. Datta {\it et al}., hep-ph/9707259.

\bibitem{shuryak} E. Shuryak and A.R. Zhitnitsky, hep-ph/9706316.

\bibitem{halperin} I. Halperin and A. Zhitnitsky, hep-ph/9704412;
hep-ph/9705251. 

\bibitem{hou} W.-S. Hou and B. Tseng, hep-ph/9705304.

\bibitem{yuan} F. Yuan and K.T. Chao, hep-ph/9706294.

\bibitem{atwood} D. Atwood and A. Soni, Phy.\ Lett.\ B{\bf405}, 150
(1997).

\bibitem{dunietz} I. Dunietz {\it et al}., hep-ph/9612421.

\bibitem{lenz} A. Lenz {\it et al}., hep-ph/9706501.

\bibitem{fleischer} R. Fleischer and T. Mannel, hep-ph/9704423.

\bibitem{london} D. London and A. Soni, Phys.\ Lett.\ B{\bf407}, 61
(1997).

\bibitem{dighe} A. Dighe {\it et al}., hep-ph/9707521.

\bibitem{atwoodtwo} D. Atwood and A. Soni, hep-ph/9706512 to appear in
PRL (Dec. 1997).

\bibitem{berkel} K. Berkelman, private communication. 

\bibitem{brodsky}
S.J. Brodsky and
P. Lepage, PRD{\bf24}, 1808 (1980); 
A. Anselm {\it et al}., hep-ph/9603444; J.M. Gerard and T. Lahana,
PLB{\bf356}, 381 (1995).

\bibitem{alitwo} A. Ali and C. Greub, PLB{\bf361}, 146 (1995);
ZPC{\bf49}, 431 (1991).


\end{thebibliography}
\end{document}